\documentclass[%
reprint,
amsmath,amssymb,
aip,
landscape,onecolumn,letterpaper
]{revtex4-1}
\usepackage{graphicx}
\usepackage{dcolumn}
\usepackage{bm}
\usepackage{afterpage}
\usepackage{amssymb}
\usepackage{graphicx}
\usepackage[version=4]{mhchem}
\usepackage[export]{adjustbox}
\usepackage[dvipsnames]{xcolor}
\usepackage{tabularx}
\usepackage{textgreek}
\usepackage{float}
\usepackage[hidelinks]{hyperref}
\usepackage{svg}
\usepackage{}
\begin{document}

\title{Femtosecond Symmetry Breaking and Coherent Relaxation of Methane Cations at the Carbon K-Edge}

\author{Enrico Ridente}
\altaffiliation{These authors contributed equally to this work.}
\affiliation{%
 Department of Chemistry, University of California, Berkeley, CA, 94720, USA
}
\affiliation{Chemical Sciences Division, Lawrence Berkeley National Laboratory, Berkeley, CA 94720, USA}
\author{Diptarka Hait} 
\altaffiliation{These authors contributed equally to this work.}
\affiliation{%
 Department of Chemistry, University of California, Berkeley, CA, 94720, USA
}
\affiliation{Chemical Sciences Division, Lawrence Berkeley National Laboratory, Berkeley, CA 94720, USA}
\affiliation{Current affiliation: Department of Chemistry, Stanford University, Stanford, CA 94305, USA}
\author{Eric A. Haugen} 
\affiliation{%
 Department of Chemistry, University of California, Berkeley, CA, 94720, USA
}
\affiliation{Chemical Sciences Division, Lawrence Berkeley National Laboratory, Berkeley, CA 94720, USA}
\author{Andrew D. Ross} 
\affiliation{%
 Department of Chemistry, University of California, Berkeley, CA, 94720, USA
}
\affiliation{Chemical Sciences Division, Lawrence Berkeley National Laboratory, Berkeley, CA 94720, USA}
\affiliation{Current affiliation: TOPTICA Photonics Inc. 1120 Pittsford Victor Road.
Pittsford, NY 14534, USA}
\author{Daniel M. Neumark} 
\affiliation{%
 Department of Chemistry, University of California, Berkeley, CA, 94720, USA
}
\affiliation{Chemical Sciences Division, Lawrence Berkeley National Laboratory, Berkeley, CA 94720, USA}
\author{Martin Head-Gordon} 
\affiliation{%
 Department of Chemistry, University of California, Berkeley, CA, 94720, USA
}
\affiliation{Chemical Sciences Division, Lawrence Berkeley National Laboratory, Berkeley, CA 94720, USA}
\author{Stephen R. Leone}
\affiliation{%
 Department of Chemistry, University of California, Berkeley, CA, 94720, USA
}
\affiliation{Chemical Sciences Division, Lawrence Berkeley National Laboratory, Berkeley, CA 94720, USA}
\affiliation{%
 Department of Physics, 
 University of California, Berkeley, CA, 94720, USA
}%

\begin{abstract}
Understanding the relaxation pathways of photoexcited molecules is essential to gain atomistic level insight into photochemistry. Herein, we perform a time-resolved study of ultrafast molecular symmetry breaking via geometric relaxation (Jahn-Teller distortion) on the methane cation. Attosecond transient absorption spectroscopy with soft X-rays at the carbon K-edge reveals that the distortion occurs within $10\pm 2$ femtoseconds after few-femtosecond strong-field ionization of methane. The distortion activates coherent oscillations in the scissoring vibrational mode of the symmetry broken cation, which are detected in the X-ray signal. These oscillations are damped within $58\pm13$ femtoseconds, as vibrational coherence is lost with the energy redistributing into lower-frequency vibrational modes. This study completely reconstructs the molecular relaxation dynamics of this prototypical example and opens new avenues for exploring complex systems.
\end{abstract}

\maketitle
\vspace{-20pt}
\section{Introduction}
\vspace{-10pt}
Chemical reactions arise from the motion of atomic nuclei. Atomic displacements can be described either in terms of fluctuations about a local minimum of energy or relaxation towards such a minimum from a nonequilibrium configuration. The latter often results from interaction with light, as photon absorption can lead to excited electronic states with minimum energy geometries quite distinct from the initial starting point. The nonequilibrium configurations arising from light-matter interaction thus can have significant surplus potential energy, which can drive chemical transformations. Therefore, intramolecular relaxation dynamics of photoexcited molecules are of fundamental photochemical interest.
 
  Jahn-Teller (JT) distortion\cite{jahn1937stability,bersuker2001modern} is a special type of relaxation mechanism that spontaneously reduces the spatial symmetry of nonlinear molecules in degenerate electronic states. Molecular geometries where multiple electronic states are isoenergetic are not stable for any of the associated states \cite{jahn1937stability} and represent a fundamental breakdown of the Born-Oppenheimer approximation \cite{longuet1961some}. It therefore becomes energetically favorable to undergo distortions that lift the degeneracy by breaking spatial symmetry. JT distortions are ubiquitous in solids \cite{persson2002structure} and gas-phase molecules \cite{koppel1988interplay}. Herein, we utilize attosecond X-ray Transient Absorption Spectroscopy (XTAS) to study symmetry-breaking of the methane cation (\ce{CH4+}) generated via vertical strong-field ionization. This unequivocally reveals the role and timescale of JT induced dynamics experimentally and provides new understanding on relaxation mechanisms in molecular systems.
 
\begin{figure}[htb!]
\includegraphics[width=\linewidth]{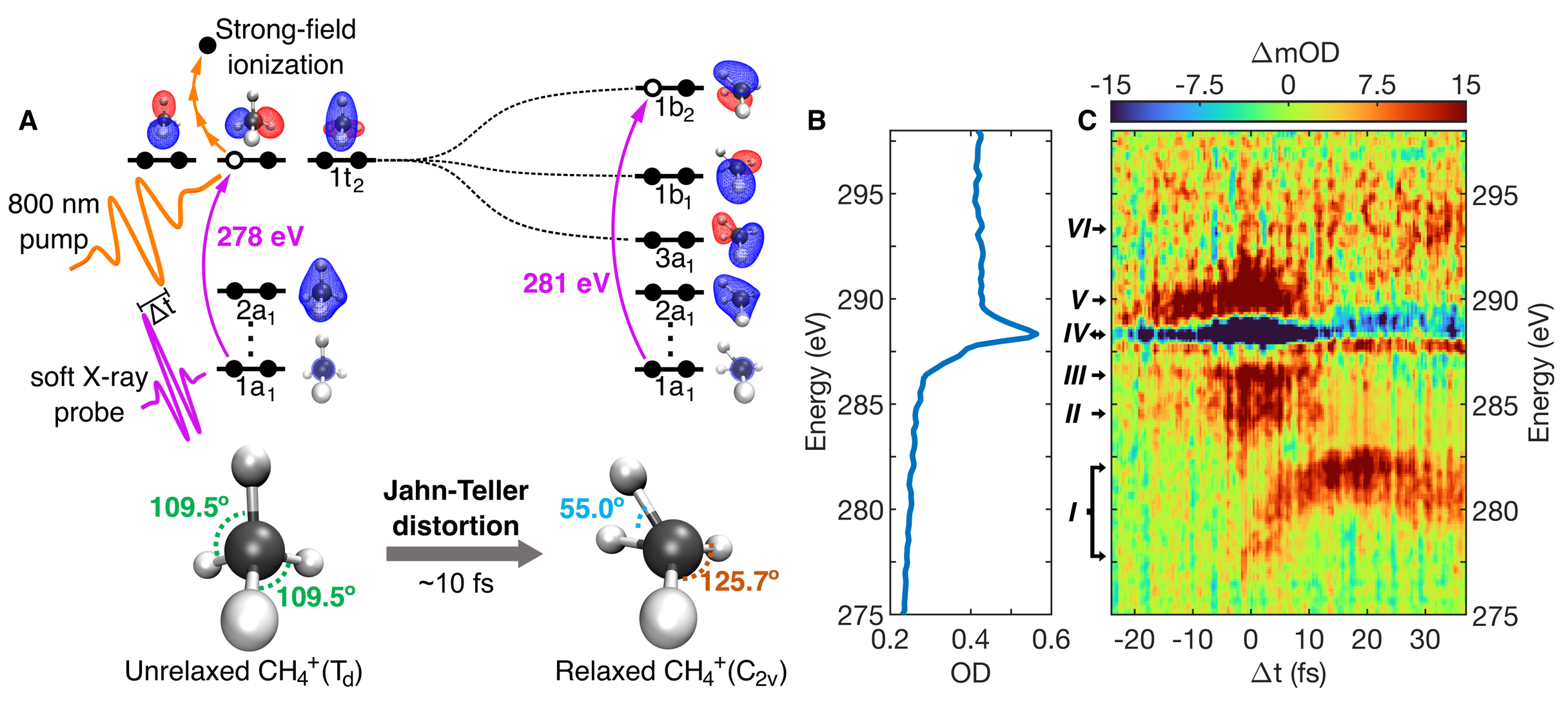}
\caption{\label{fig:CH4Struct} (A) Schematic of the pump-probe process. Pump-induced strong-field ionization produces \ce{T_d} symmetry \ce{CH4+}, which undergoes Jahn-Teller distortion towards a \ce{C_{2v}} minimum. Molecular orbitals for both structures are also shown (energies not to scale). The 1a$_1$ orbital is the C 1s core-level while the remainder are \ce{\sigma_{C-H}} bonding orbitals. The dynamics are mapped by the 1s$\to$SOMO probe transition induced by a time-delayed ($\Delta$t) soft X-ray probe. (B) Ground state carbon K-edge X-ray absorption spectrum for \ce{CH4}. (C) Transient X-ray absorption at $\Delta$t $<$ 35 fs (negative time indicates probe preceding pump). The low energy signal (278-282 eV) corresponds to the 1s$\to$SOMO probe transition, which indicates that the \ce{C_{2v}} minimum is reached by $\approx$10 fs.\vspace{-15pt}}
\end{figure}

 \ce{CH4+} is a classic system in which JT distortions occur \cite{potts1972photoelectron,knight1984experimental,signorell1999jahn,gonccalves2021ultrafast}. The process starts with \ce{CH4}, which is the smallest stable molecule with tetrahedral (\ce{T_d}) geometry. The equilibrium C–H bond distances are 1.087 {\AA} \cite{hirota1979anharmonic}, and all the H–C–H bond angles are $\approx$ 109.5$^\circ$ due to \ce{T_d} symmetry. The ground state molecular orbitals (MOs) are shown in Fig. \ref{fig:CH4Struct}A and the electronic configuration of neutral \ce{CH4} is 1a$_1^2$2a$_1^2$1t$_2^6$. The electronic ground state configuration of \ce{CH4+} at the \ce{T_d} geometry is 1a$_1^2$2a$_1^2$1t$_2^5$, which is triply degenerate as each of the three 1t$_2$ orbitals is equally likely to be singly occupied. Therefore, \ce{CH4+} undergoes JT distortion away from the \ce{T_d} geometry to a lower symmetry \ce{C_{2v}} form \cite{knight1984experimental,signorell1999first,signorell2000pfi,worner2006jahn,worner2007jahn}. This structure is computed to have two long (1.187 {\AA}) and two short (1.083 {\AA}) C–H bonds (as shown in Fig. \ref{fig:CH4Struct}A). The angle formed by the long C–H bonds is 55.0$^\circ$ while the short bonds form an angle of 125.7$^\circ$, representing significant deviations from the initial tetrahedral geometry. These distortions lower the energy of the doubly occupied 3a$_1$ and 1b$_1$ MOs (Fig. \ref{fig:CH4Struct}A), but also destabilize the 1b$_2$ singly occupied MO (SOMO). The electronic ground state of \ce{CH4+} therefore is $^2$B$_2$\cite{signorell2000pfi}.

Molecular JT distorted forms, and \ce{CH4+} in particular, have been extensively studied\cite{potts1972photoelectron,rabalais1971jahn}. Nevertheless, time-resolving the JT distortion in \ce{CH4+} remained an open challenge \cite{knight1984experimental,worner2006jahn}, due to the ultrafast nature of the process. JT distorted species are conventionally observed in the rovibrational spectrum of photoelectron experiments \cite{potts1972photoelectron,baltzer1995experimental}, but such measurements lack the temporal resolution to obtain the femtosecond timescale dynamics of symmetry breaking. Baker et al. \cite{baker2006probing} utilized attosecond resolution high-harmonic emission spectroscopy to report on the onset of the JT distortions in \ce{CH4+} and deuterated \ce{CD4+} up to the first 1.6 fs. The nuclear motion in those experiments, however, cannot be reconstructed at longer times, precluding a complete analysis of the JT relaxation process and subsequent coherent motion. Coulomb explosion experiments by Li et al. \cite{li2021ultrafast} probed the dynamics of \ce{CH4+} by recording the photofragments after interaction with two time-delayed strong-field 800 nm pulses, but temporal resolution was limited by the 25 fs pulses used as pump and probe. Furthermore, their use of a multi-cycle pump pulse led to several additional photoproducts from higher energy fragmentation pathways that compete with JT distortions. Indeed, it has been shown that the use of shorter, few-cycle 800 nm pulses increases the relative amount of \ce{CH4+} by suppressing additional product channels \cite{varvarezos2021ionization}.
 
  XTAS, based on attosecond and few-femtosecond duration soft X-ray pulses generated via high-harmonic generation \cite{barreau2020efficient}, has been successfully used to study ultrafast molecular relaxation processes with high structural and temporal resolution \cite{geneaux2019transient,pertot2017time}. XTAS at the carbon K-edge is therefore an ideal platform to observe few-femtosecond timescale dynamics\cite{zinchenko2021sub,ross2022jahn} like those associated with the JT distortion of \ce{CH4+}. The X-ray probe excites C 1s electrons to unoccupied levels such as the SOMO or completely unoccupied antibonding/Rydberg levels. In particular, the dipole-allowed 1s$\to$SOMO signal in \ce{CH4+} is expected to be energetically well-resolved from other features. Geometric changes as a result of JT distortion will strongly affect the SOMO energy, which can be traced by XTAS with few-femtosecond time resolution.
  
 In this work, we report a joint experimental and theoretical study of the symmetry-breaking JT dynamics of \ce{CH4+}. The cations are produced from neutral methane via abrupt few-femtosecond strong-field ionization (SFI) of \ce{CH4} with an 800 nm, few-cycle pump pulse generated by a table-top Ti:Sapphire laser. The induced dynamics are then probed with XTAS using high-harmonic generated soft X-ray pulses at the C K-edge obtained with a 1300 nm source \cite{barreau2020efficient}. The nonperturbative nature of SFI leads to an ionization window that is temporally much narrower than the 5 fs width of the pump pulse \cite{smirnova2006coulomb}. We observe a significant energy shift in the XTAS signal immediately upon ionization due to JT distortion. The \ce{C_{2v}} minimum geometry is attained within $\approx$10 fs, which is followed by two coherent oscillations in the signal that reveal large amplitude scissoring motion. These coherent oscillations are damped out by $\approx$ 60 fs, indicating vibrational dephasing and eventual decoherence. The behavior of fully deuterated methane cations (\ce{CD4+}) is also investigated to understand the effect of substituted masses on the JT dynamics.
\section{Methods}\vspace{-10pt}
\subsection{Experiment}\vspace{-10pt}
 The experimental setup used to perform the measurements reported in the main text was described in depth by Barreau et al.\cite{barreau2020efficient} and is briefly summarized here. 800 nm pulses generated by a Ti:Sapphire oscillator are amplified by a multi-pass and single-pass Ti:Sapphire chirped-pulse amplifier that operates at 1 kHz. After amplification and temporal compression, the pulses have an energy of 12 mJ and a pulse duration of $\approx$ 36 fs. A beamsplitter is used to separate the pump (800 nm) and probe (soft X-ray) arms. The probe arm consists of an 85\% reflection of the radiation off the beamsplitter to an optical parametric amplifier. Here the central wavelength is down converted to 1300 nm pulses with an energy of 2.5 mJ. This beam is subsequently coupled through a stretched hollow-core fiber filled with 0.5 bar of Ar. After the fiber, 1.25 mJ, 11 fs-short pulses are produced via temporal compression with chirped mirrors and SF11 wedges. The 1300 nm pulses are then used to generate, via HHG, soft X-ray pulses in a semi-infinite gas cell filled with $\approx$2.5 bar of He. After the gas cell, a 100 $\mu$m titanium filter is used to separate the 1300 nm light from the soft X-ray that is transmitted through the filter. The soft X-ray pulses have a spectrum that spans up to 375 eV, thus easily reaching the carbon K-edge (270-300 eV). Finally, the soft X-ray beam is focused by a toroidal mirror on the sample cell, where it is recombined with the time-delayed 800 nm pump. The latter is obtained from the residual 15\% of the 800 nm power transmitted through the beamsplitter after the Titanium:Sapphire chirped-pulse amplifier. This beam is sent to a second flexible hollow-core fiber, filled with 0.25 bar of Ar and compressed using chirped mirrors, ammonium dihydogen phosphate (ADP) and fused silica. The 800 nm pulses thus obtained have an energy of $\approx$ 150 $\mu$J and a pulse duration of $\approx$ 5 fs (two optical cycles). 

 Finally, the pump is sent into the vacuum chamber where it is focused on the flowing sample gas cell using a mirror with a focal length of 300 mm, resulting in a 65 $\mu$m diameter focus and a maximum intensity of $\approx$ 3.5$\times$10$^{14}$ W/cm$^2$. The energy of the pump pulse is tuned with a broadband half-waveplate and a polarizer to actively control the pump intensity from 1$\times$10$^{14}$ to 3.5$\times$10$^{14}$ W/cm$^2$, thereby allowing a fine tunability of the pump power to achieve the minimum energy necessary to ionize \ce{CH4} while curtailing the emergence of high energy channels from higher pump intensities. It is found that an intensity of $\approx$ 3$\times$10$^{14}$ W/cm$^2$ results in the clearest signal of the Jahn-Teller distortion and corresponds with the experimental results reported in this work, as discussed in the supporting information (SI). Previous investigations\cite{wu2007ionization} confirm that the major photoproduct for an 800 nm pump pulse of similar intensity and pulse duration is \ce{CH4+} . The pump is spatially overlapped with the soft X-ray beam on the sample gas cell using an annular mirror, thus, minimizing temporal broadening due to a noncollinear geometry. The sample gas cell consists of a hollow cylinder with a 4-mm diameter, through which a 200 $\mu$m hole is drilled. The cell is operated with a \ce{CH4} backing pressure of 30 mbar. After the sample cell, the pump is blocked with a 100 nm titanium filter, allowing most of the soft X-rays to be transmitted while reflecting the 800 nm beam. The soft X-rays are subsequently diffracted by a grating and imaged on a CCD camera. The system provides a spectral resolution of 0.3 eV. The autoionization of the Ar L$_{2,3}$ lines provides a measure of the cross-correlation between pump and probe. This is measured to be 9$\pm$1 fs (see SI). This value is a measure of the convolution of the pump and probe pulse durations in addition to the timing jitter between the two. It is expected that with the few-cycle pump pulse that strong-field ionization may only occur at the most intense peaks of the electric field, resulting in a shorter time window for ionization\cite{smirnova2006coulomb}. The measurements have been processed and obtained using the same filtering procedure employed by Ott et al. \cite{ott2014reconstruction}.
 \vspace{-10pt}
\subsection{Theory}\vspace{-10pt}
All quantum chemical calculations utilized the Q-Chem 5 package\cite{epifanovsky2021software}. Quasiclassical ab initio molecular dynamics (AIMD) calculations\cite{paranjothy2013direct} on \ce{CH4+} were performed at the $\omega$B97M-V\cite{wB97MV}/pcseg-2\cite{jensen2014unifying} level. The initial positions and velocities of the nuclei were obtained via sampling the Wigner quasiprobability distribution\cite{wigner1932quantum} corresponding to the neutral \ce{CH4} ground state harmonic wavefunction (utilizing the geometry and frequencies computed at the same $\omega$B97M-V/pcseg-2 level). The dynamics therefore simulate the behavior after vertical ionization within the Franck Condon approximation. The timestep for the dynamics was 5 atomic units (i.e., $\approx$ 0.1209 fs), and the trajectories were run for 1000 steps ($\approx$ 120.9 fs). All trajectories were run under constant energy conditions (i.e., no thermostat), on the electronic ground state of \ce{CH4+}. No non-adiabatic effects were included in the calculations, nor was an explicit treatment for nuclear quantum effects made. The trajectories were run without any external fields, and thus any post-ionization effects of the pump pulse were not explicitly treated. 256 trajectories corresponding to distinct initial positions and momenta were run for \ce{CH4+}. Five initial conditions out of 256 had potential energy $>$ 1.5 eV, and three of those showed bond breaking behavior to form \ce{CH3+} and H. As no significant signal corresponding to \ce{CH3+} was experimentally observed outside of very high pump intensities (see SI), these three trajectories were discarded for analyses of \ce{CH4+} (which therefore utilized a total of 253 trajectories).
 The dynamics of \ce{CD4+} were simulated similarly, utilizing a set of 256 distinct positions and momenta sampled from the Wigner quasiprobability distribution corresponding to the ground state harmonic wavefunction of neutral \ce{CD4}.  The timestep for \ce{CD4+} was 10 atomic units ($\approx$ 0.2418 fs), and a total of 440 timesteps ($\approx$ 105.6 fs) were run. None of the 256 trajectories utilized for \ce{CD4+} showed dissociative behavior (likely due to lower zero-point energy) and thus all were used for subsequent analyses.
 
 The geometries constituting the AIMD trajectories were used for computing the X-ray absorption spectrum with orbital-optimized DFT (OO-DFT \cite{hait2021orbital,hait2020highly,hait2020accurate}). Specifically, the geometries after every 5 timesteps (25 atomic units or $\approx$0.6047 fs) of the 253 nondissociative \ce{CH4+} trajectories, until $\approx$103.4 fs, were selected. These geometries (171 for each trajectory) are ‘snapshots’ of the molecule over the time-evolution and can be utilized to compute core-level spectra. The 1s$\to$SOMO excitation for these snapshots were computed using $\Delta$SCF\cite{besley2009self} with the SCAN\cite{SCAN} functional, using the procedure described in Ref \citenum{hait2020accurate}. The square gradient minimization (SGM \cite{hait2020excited})  approach was used for restricted open-shell optimizations of core-hole states, and the initial maximum overlap method (IMOM\cite{barca2018simple}) for spin-unrestricted calculations on such states. The pcX-1 basis was utilized on the C atom for the core-level spectrum calculations, as this basis is optimized for core-level calculations\cite{ambroise2018probing}, while the pcseg-1 basis was used on the H atoms. Scalar relativistic effects in the core-level calculations were accounted for with the spin-free one electron exact two component (SFX2C-1e) approach\cite{saue2011relativistic,cunha2022relativistic}. The resulting excitation energies and oscillator strengths were used to compute theoretical XTAS. Snapshots for \ce{CD4+} were selected at intervals of 5 timesteps (50 atomic units or $\approx$1.209 fs), until $\approx$103.4 fs.
 
\begin{figure}[thb!]
\includegraphics[width=\linewidth]{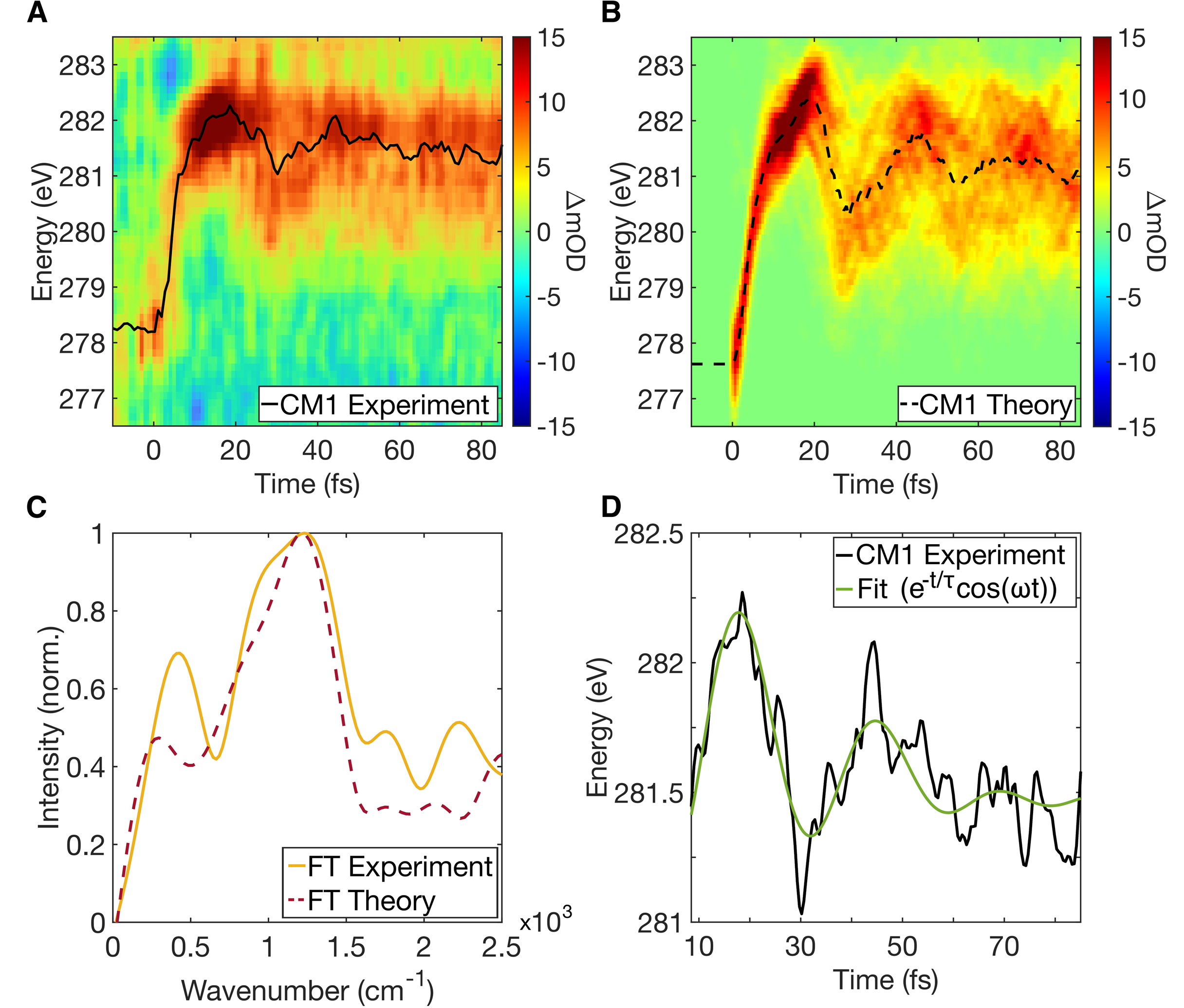}
\caption{\label{fig:CH4spectrum}  (A) Experimental XTAS for the 1s$\to$SOMO transition of \ce{CH4+}, with the first central moment (CM1) shown in black. (B) Theoretical XTAS for the same excitation. The CM1 is computed only over regions with absorbance $>$15$\%$ of the peak, to account for experimental signal-to-noise. (C) Fourier transform of the CM1 from experiment (solid line) and theory (dashed line). The theoretical intensities have been uniformly scaled to match experiment for peak absorbance. (D) Experimental CM1 fit with $-\exp\left(-\frac{t}{\tau}\right)\cos\omega t$, with $\omega$ = 1200 cm$^{-1}$, indicating a damping lifetime of $\tau$ = 58$\pm$13 fs for the vibrational dephasing.
\vspace{-15pt}}
\end{figure}

\section{Results and Discussion}\vspace{-10pt}
\subsection{General features of XTAS signal}\vspace{-10pt}
\ce{CH4} has a simple ground state X-ray absorption spectrum (as shown in Fig. \ref{fig:CH4Struct}B). There is only one prominent peak (288 eV), which arises from the 1s$\to$3p Rydberg excitation\cite{schirmer1993k}. No other noteworthy pre-edge features are observed and 1s ionization occurs at $\approx$290.8 eV \cite{pireaux1976core}.
 Fig. \ref{fig:CH4Struct}C shows the experimental carbon K-edge XTAS spectrum for \ce{CH4+} up to 35 fs after the pump pulse abruptly ionizes the molecule at t = 0.  The negative (blue) transient signal (feature IV) corresponds to the depletion of neutral \ce{CH4}. At t = 0, other prominent features are the positive (red) signals at 278-282 eV (I), 284 eV (II), 287 eV (III) and 290-292 eV (V). The broad features II, III, and V have significant temporal overlap with the pump pulse and can be attributed to the Stark effect of the pump pulse on core-excitation energies of \ce{CH4} (see SI). After t = 0, a positive feature at $\approx$287.5 eV can be observed. This arises because of Raman activation of the symmetric stretch vibrational mode of \ce{CH4} by the pump pulse (see SI), similar to the behavior observed in other molecules \cite{ferre2014high,chang2022conical}.
  Feature I can be assigned to \ce{CH4+} based on OO-DFT calculations that reveal this feature corresponds to the 1s$\to$SOMO excitation of nonequilibrium, \ce{T_d} \ce{CH4+}. OO-DFT indicates that other core-level excitations of \ce{CH4+} (1s$\to \sigma^*$/Rydberg levels) are above 288 eV in energy (see SI). Feature VI in Fig. \ref{fig:CH4Struct}C corresponds to such excitations and can be observed at long times. However, feature VI does not show discernible time-evolution, potentially due to overlap with 1s core-ionization of \ce{CH4}. Conversely, feature I is well separated from all the other features and is particularly sensitive to changes in molecular geometry as the SOMO is of \ce{\sigma_{C-H}}  character. The time-evolution of this signal is the clearest reporter of the dynamics of \ce{CH4+} and the analysis below therefore focuses on it.

\vspace{-10pt}
\subsection{Relaxation dynamics of \ce{CH4+}}\vspace{-10pt}
The long-time experimental XTAS of the JT feature corresponding to the 1s$\to$SOMO transition is given in Fig. \ref{fig:CH4spectrum}A. The energetic average (henceforth abbreviated as CM1, for first central moment) of the differential absorption ($\Delta$mOD) signal (solid black line) shows three main characteristics. A rapid blueshift in energy from 278 eV (at t $\approx$ 0) to $\approx$ 282 eV (at t $\approx$ 18 fs) is followed by damped oscillations until t $\approx$ 60 fs and subsequently an almost time-independent signal between 281-281.5 eV. The width of the spectral feature increases significantly starting around t $\approx$ 10 fs, leading to a very broad signal at longer times. 

 We interpret the behavior of this signal via AIMD trajectories on \ce{CH4+}. Fig. \ref{fig:CH4spectrum}B shows the XTAS spectrum computed via OO-DFT from 253 nondissociative AIMD trajectory geometries, revealing good agreement with the experimental results. This comparison indicates that the trajectories underlying the spectrum are a good reporter of the molecular dynamics under the experimental conditions. The timescale for the JT process can be directly estimated by the time taken by the CM1 to attain the 281.5 eV value that OO-DFT estimates as the upper bound for the 1s$\to$SOMO excitation of \ce{CH4+} at the \ce{C_{2v}} equilibrium geometry (see SI). This maximum is measured at 10$\pm$2 fs from experiment, and calculated to be 9.6$\pm$0.4 fs from theory, confirming the rapidity of JT distortion in \ce{CH4+} relative to typical vibrational timescales.
 
  In general, the time-evolution of the signal corresponds to atomic motions associated with the relaxation process, with the oscillatory patterns suggesting involvement of vibrational modes of \ce{CH4+}. This evolution can be analyzed further via a Fourier transform (FT) of the CM1 position from both theory and experiment (Fig.  \ref{fig:CH4spectrum}C). Even though the FT features are broad owing to the rapid decay in the oscillation amplitude, it is possible to identify critical frequencies. The most intense peak in the FT is at $\approx$ 1200 cm$^{-1}$, a frequency that corresponds to a computed normal mode of the \ce{C_{2v}} minimum associated with scissoring about the smallest bond angle, i.e. the angle between the two long C-H bonds. However, caution must be taken in interpreting the FT features in terms of the fundamental frequencies of the \ce{C_{2v}} minimum; the \ce{CH4+} ground state surface has twelve distinct \ce{C_{2v}} minima\cite{paddon1985dynamic} and multiple seams corresponding to electronic state degeneracies, resulting in a highly anharmonic potential energy surface (PES). The FT is nonetheless an indication of several molecular motions that affect the signal and the associated timescales. The damping rate for the 1200 cm$^{-1}$ frequency can also be estimated by fitting to the time-domain experimental CM1 (Fig. \ref{fig:CH4spectrum}D), revealing a lifetime of 58$\pm$13 fs for the oscillations.

\begin{figure}[thb!]
\includegraphics[width=\linewidth]{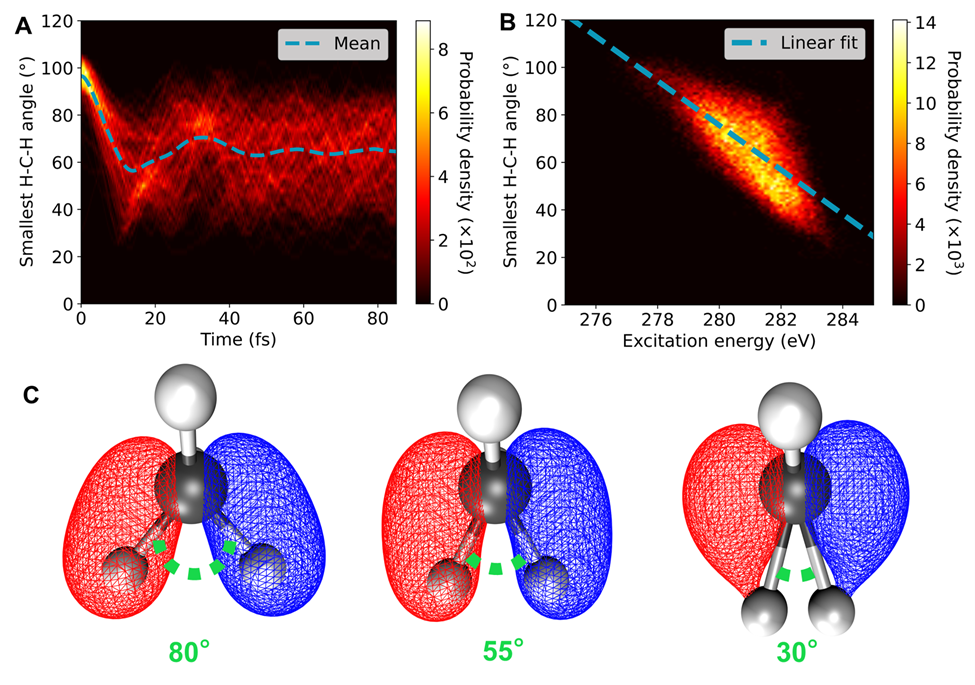}
\caption{\label{fig:theory} (A) Time-evolution of the smallest bond angle over the trajectories. (B) Correlation between the computed 1s$\to$SOMO excitation energies vs the smallest bond angle of the corresponding structures. (C) Evolution of the SOMO with change in the smallest bond angle.\vspace{-15pt}}
\end{figure}

We utilize AIMD trajectories to uncover the origins of the signal oscillations in the X-ray spectra. The trajectories indicate that the C-H bond lengths oscillate on a timescale roughly twice as fast as the principal oscillation in the CM1 (see SI). The angular oscillations are slower than the stretches, with Fig. \ref{fig:theory}A showing that the mean of the smallest molecular bond angle over all trajectories undergoes a damped oscillatory motion on the same timescale as the predicted and observed XTAS signal CM1 (Figs. \ref{fig:CH4spectrum}A and \ref{fig:CH4spectrum}B). The computed bond angle distribution also broadens rapidly over time after t$\approx$10 fs, similar to the XTAS signal. A direct correlation between the observed signal and the smallest bond angle is revealed by Fig. \ref{fig:theory}B, which plots the smallest bond angle of the trajectory geometries utilized for Fig. \ref{fig:CH4spectrum}B against the computed 1s$\to$SOMO excitation energies for those geometries. It is evident that a simple linear model can capture much of the relationship between the two quantities. The excitation energy has weaker correlation with other bond angles and bond lengths (see SI). Theory therefore indicates that the most important contribution to the observed time-evolution of the XTAS absorption energy is from the dynamics of the smallest bond angle, to the extent that the signal can be interpreted in terms of a single parameter.

This connection can be understood via a simple orbital model (Fig. \ref{fig:theory}C). The SOMO at the \ce{C_{2v}} symmetry minimum geometry of \ce{CH4+} is the bonding orbital arising from mixing between a C 2p orbital and a symmetry-adapted linear combination (SALC) of the 1s orbitals corresponding to H atoms in the long bonds. This SALC has antibonding character, as the two H 1s orbitals have opposite phases. For small bond angles, the H 1s SALC has poorer overlap with the C 2p level as it gets closer to the nodal plane of the latter. This leads to a weaker interaction and therefore lowers the mixing between the C and H centered orbitals. Furthermore, smaller angles lead to decreased H-H distance, elevating the energy of the H 1s SALC due to the local antibonding character. The resulting \ce{\sigma_{C-H}} bonding orbital therefore has greater nonbonding (pure C 2p) character as the angle decreases, made evident in the 30$^\circ$ angle in Fig. \ref{fig:theory}C. Conversely, larger bond angles lead to a more stabilized SOMO with greater contribution from H orbitals. This picture is consistent with the observed increase in the 1s→SOMO X-ray probe excitation energy with decreasing bond angle shown in Fig. \ref{fig:theory}B. The X-ray oscillator strength also increases with a decrease in bond angle (see SI), highlighting the increase in C 2p character of the SOMO. Therefore, the XTAS signal reveals the extent to which the SOMO loses C-H bonding character during the relaxation process.

  It is thus apparent that the \ce{T_d\to C_{2v}} JT distortion activates the scissoring mode about the smallest bond angle in the \ce{C_{2v}} minimum, which is the most evident feature in the X-ray spectra. For a perfectly harmonic PES, the excess energy accumulated in this mode would remain undissipated therein, leading to undamped oscillation of the XTAS signal and geometric parameters about the minimum before radiative relaxation to the vibrational ground state. However, the PES of \ce{CH4+} is highly anharmonic (see discussion above), and the surplus energy spreads out to all other modes. This redistribution is observed via damping in the oscillations for both the experimental XTAS signal CM1 and the mean geometric parameters from the AIMD trajectories (Figs. \ref{fig:CH4spectrum} and \ref{fig:theory}). In addition, considerable broadening of the XTAS signal is observed after the initial 10 fs, which is mirrored by an increase in the width of the probability distributions for the geometric parameters. Fig. \ref{fig:CH4spectrum}D indicates that the experimental XTAS CM1 oscillations have a damping lifetime of 58 fs, while oscillations in parameters computed from AIMD trajectories are mostly damped out within 60 fs (see SI). It therefore appears that a large proportion of energy is transferred out of the JT activated scissoring mode to other internal degrees of freedom within this timescale, constituting an ultrafast example of intramolecular vibrational energy redistribution (IVR). This process can also be described in terms of dephasing of the scissoring mode (i.e., relaxation towards `thermal equilibrium’), which is ultimately reflected in the decoherence of the XTAS signal.

\begin{figure}[htb!]
\includegraphics[width=0.6\linewidth]{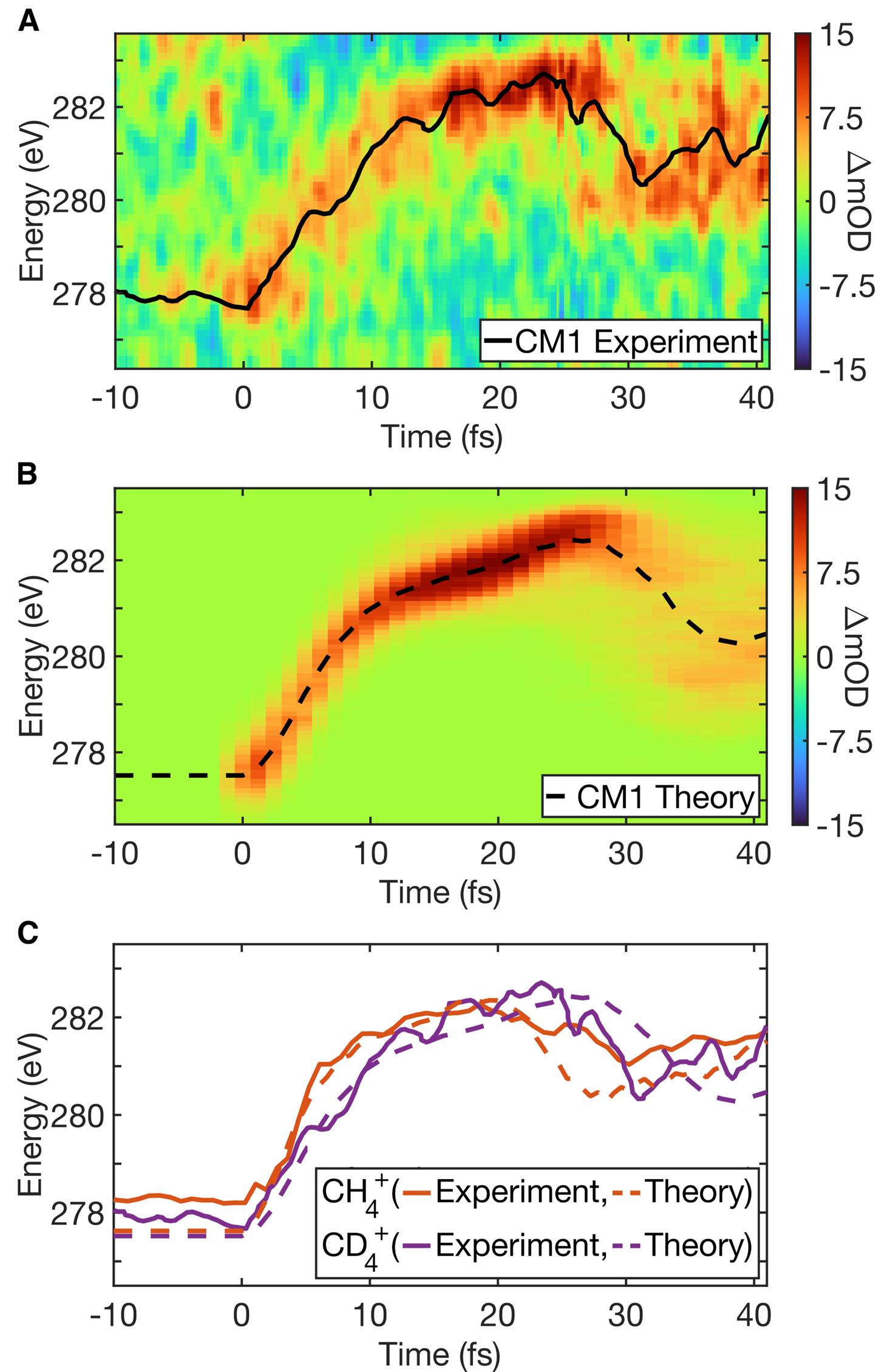}
\caption{\label{fig:CD4spectrum} (A) Experimental XTAS of the 1s$\to$SOMO transition of \ce{CD4+}, with the CM1 in black. (B) Theoretical XTAS for the same excitation. (C) Comparison of the CM1 for \ce{CH4+} (orange) and \ce{CD4+} (violet) from experiment (solid line) and theory (dashed line). The JT distortion is slower for \ce{CD4+} than \ce{CH4+} by a factor of $\approx$1.4.}
\end{figure}

Comparison between experiment and theory therefore reveals that the XTAS signal is reporting the nuclear motion over the duration of the relaxation process. It is important to consider that SFI experiments can exhibit electronic relaxation dynamics without significant nuclear motion, involving highly excited Rydberg or cationic states\cite{scherzer1994long,kobayashi2020coherent}. To prove that the observed XTAS signal dynamics solely arise from nuclear (rather than electronic) relaxation, we performed experiments and computations for the dynamics following SFI of \ce{CD4}. The greater mass of the deuterium isotope should slow the scissoring motion, leading to slower time-evolution of the signal. This effect is experimentally observed (Fig. \ref{fig:CD4spectrum}A), unambiguously revealing that the signal is reporting on nuclear dynamics. Fig. \ref{fig:CD4spectrum}B shows that the computed XTAS spectrum from OO-DFT on 256 AIMD trajectories of \ce{CD4+} is also consistent with the experiment, further validating the ability of the computed trajectories to successfully simulate the experimental molecular dynamics. Fig. \ref{fig:CD4spectrum}C shows the time-evolution of the CM1 for both \ce{CH4+} and \ce{CD4+} from experiment and theory. The time taken by the CM1 to reach the 281.5 eV value associated with the \ce{C_{2v}} cation geometry is longer for \ce{CD4+} than for \ce{CH4+}, and the effect of deuteration can be gauged by the ratio (\ce{CD4+}/\ce{CH4+}) of these times. This is measured to be 1.4$\pm$0.3 from experiment and calculated to be 1.47$\pm$0.07 from theory (see SI). The value thus calculated is close to the ratio between the scissoring frequencies for the smallest bond angle in \ce{CH4+} and \ce{CD4+} (computed to be 1.34). However, it is important to note that all the normal modes of \ce{C_{2v}} \ce{CH4+} have similar frequency ratios upon isotopic substitution (see SI). We also note that Baker et al. \cite{baker2006probing} reported that dynamics within the first 1.6 fs of ionization was a factor of 2-3 slower for \ce{CD4+} vs \ce{CH4+}. This behavior at very short times results from the decay of the autocorrelation of the nuclear wavefunction \cite{baker2006probing,gonccalves2021ultrafast}, which is distinct from the longer time dynamics reported here involving significant atomic displacements, therefore representing almost nonoverlapping nuclear wavefunctions.

\section{Conclusions}\vspace{-10pt}
  \ce{CH4+} was prepared from strong-field ionization of \ce{CH4} and probed with transient X-ray absorption spectroscopy near the carbon K-edge with few-femtosecond time resolution. Evolution of the excitation from the C 1s level to the valence hole reveals the dynamics of JT symmetry-breaking away from the parent tetrahedral geometry, as well as subsequent coherent motion and dissipation of released energy out of active modes. All three of these aspects of intramolecular relaxation have been successfully observed and analyzed. The combination of experiment and theory reveals that the molecule first reaches the JT distorted form within 10$\pm$2 fs after ionization. This distortion involves reduction of a H-C-H bond angle from 109.5$^\circ$ towards 55$^\circ$, which is directly reported by a blue-shift in the X-ray absorption signal. The JT dynamics are found to be 1.4$\times$ slower in deuterated methane on account of the larger substituent mass, proving that the observed dynamics arise from nuclear motions. The energy released by the JT distortion drives a few coherent oscillations in the activated modes before being distributed over other molecular internal degrees of freedom, leading to damping of the oscillations within 60 fs of ionization. We note that the observed behavior for \ce{CH4+} is distinct from previous XTAS studies \cite{pertot2017time,ross2022jahn} on the dynamics of \ce{CF4+} and \ce{CCl4+}, as those species are highly unstable against bond dissociation. \ce{CF4+} has not been experimentally detected to date \cite{pertot2017time}; metastable \ce{CCl4+} has been previously observed \cite{ross2022jahn} but signals from the intramolecular relaxation pathways were unable to be disentangled from bond breaking. The subsequent coherence and dissipation of energy from the JT activated scissoring mode to other internal degrees of freedom is only observed in \ce{CH4+}. This work thus opens the door to studies on how ultrafast vibrational coherence influences the redistribution of excess energy in more complex systems.

\section{Acknowledgements}\vspace{-10pt}
This work is funded by the DOE Office of Science, Basic Energy Science (BES) Program, Chemical Sciences, Geosciences and Biosciences Division under Contract no. DE-AC02-05CH11231, through the Gas Phase Chemical Physics program (ER, EAH, ADR, DMN, SRL) and Atomic, Molecular, and Optical Sciences program (DH and MHG). The instrument was built with funds from the National Science Foundation through NSF MRI 1624322 and matching funds from the Lawrence Berkeley National Laboratory, the College of Chemistry, the Department of Physics, and the Vice Chancellor for Research at UC Berkeley. This research used resources of the National Energy Research Scientific Computing Center, a DOE Office of Science User Facility supported by the Office of Science of the U.S. Department of Energy under Contract No. DE-AC02-05CH11231 using NERSC award BES-ERCAP0020263. ADR was additionally funded by the U.S. Department of Energy, Office of Science, Office of Basic Energy Sciences, Materials Sciences and Engineering Division, under Contract No. DE-AC02-05-CH11231 within the Physical Chemistry of Inorganic Nanostructures Program (KC3103), by the W. M. Keck Foundation Grant No. 042982, and by the U.S. Army Research Office (ARO) under Grant No. W911NF-20-1- 0127.
\section{Author contributions}\vspace{-10pt}
\noindent Experimental investigation: ER, EAH, ADR
\\Theoretical investigation: DH
\\Experiments supervision: SRL, DMN
\\Theory supervision: MHG
\\Writing – original draft: ER, DH, EAH
\\Writing – review \& editing: ER, DH, EAH, ADR, MHG, DMN, SRL
\section{Conflicts of Interest}\vspace{-10pt}
The electronic structure calculations were performed in Q-Chem, which is partially owned by MHG. The other authors declare no conflict of interests.

\section{References}\vspace{-10pt}
\bibliography{references}
\end{document}